\documentclass[traditabstract,longauth]{aa}
\usepackage{hyperref}
\hypersetup{
    bookmarks=true,                 % show bookmarks bar?
    unicode=false,                  % non-Latin characters in Acrobat bookmarks
    pdftoolbar=true,                % show Acrobat toolbar?
    pdfmenubar=true,                % show Acrobat menu?
    pdffitwindow=true,              % page fit to window when opened
    pdftitle={Aperture Effects},    % title
    pdfauthor={Author},             % author
    pdfsubject={Subject},           % subject of the document
    pdfnewwindow=true,              % links in new window
    pdfkeywords={keywords},         % list of keywords
    colorlinks=true,                % false: boxed links; true: colored links
    linkcolor=blue,                 % color of internal links
    citecolor=blue,                 % color of links to bibliography
    filecolor=blue,                 % color of file links
    urlcolor=blue                   % color of external links
}

\usepackage{lscape}
\usepackage[section]{placeins}
\usepackage{wasysym}
\usepackage{graphics}
\usepackage{graphicx}
\usepackage{txfonts}
\usepackage{color}
\usepackage{times}
\usepackage{mathtools}
\usepackage[T1]{fontenc}

\newfont{\vcap}{cmssdc10 scaled 1000}
\newfont{\lcap}{cmssdc10 scaled 1100}
\newfont{\nlx}{cmssdc10 scaled 900}
\newfont{\xnlx}{cmssdc10 scaled 800}
\newfont{\hvss}{cmssdc10 scaled 1540}

\def\rr{{\sl R}$^{\star}$}
\def\ha{H$\alpha$}
\def\ewha{EW(\ha)}

\def\lo3hb{$\log$([O\,{\sc iii}]/H$\beta$)}
\def\ln2ha{$\log$([N\,{\sc ii}]/H$\alpha$)}
\def\tauha{$\tau$}
\def\tauha_ext{$\tau$}

\def\rr{{\sl R}$^{\star}$}
\def\eqan{\begin{equation}}
\def\eqen{\end{equation}}

\def\h1{\ion{H}{i}}
\def\h2{\ion{H}{ii}}
\def\reff{{\it R$_{\rm eff}$}}

\def\rr{{\sl R}$^{\star}$}

\def\ha{H$\alpha$}
\def\ewha{EW(\ha)}

\def\o5007{[O {\sc iii}] $\lambda$5007}

\def\n2ha{[N\,{\sc ii}]/H$\alpha$}
\def\ln2ha{$\log$([N\,{\sc ii}]/H$\alpha$)}
\def\tn2ha{[N\,{\sc ii}]${\scriptstyle 6584}$/H$\alpha$}
\def\tln2ha{$\log$([N\,{\sc ii}]${\scriptstyle 6584}$/H$\alpha$)}
\def\o3hb{[O\,{\sc iii}\,]/H$\beta$}
\def\lo3hb{$\log$([O\,{\sc iii}]/H$\beta$)}
\def\to3hb{[O\,{\sc iii}]${\scriptstyle 5007}$/H$\beta$}
\def\tlo3hb{$\log$([O\,{\sc iii}]${\scriptstyle 5007}$/H$\beta$)}
\def\e16{$10^{-16}~{\rm erg\,s^{-1}\,cm^{-2}}$}
\def\e17{$10^{-17}~{\rm erg\,s^{-1}\,cm^{-2}}$}
\def\tauha{$\tau$}
\def\tauha_ext{$\tau$}
\def\ha{H$\alpha$}
\def\ewha{EW(\ha)}

\def\tauha{$\tau$}
\def\tauha_ext{$\tau$}

\def\mstar{${\cal M}_{\star}$}

\def\porto3d{\sc Porto3D\rm}
\def\p3d{\sc Porto3D\rm}
\def\P3D{\sc Porto3D\rm}

\newcommand\btab[5]{\begin{table}[#1]\label{#3}{\parbox{#4}{\caption{#2}}\rule[-0.5ex]{0cm}{0.5ex} }
\begin{tabular*}{#4}{#5} \label{#3} }
\newcommand{\etab}[3]{
\end{tabular*}

\vspace*{#1}
\begin{flushleft}
\parbox{#2}{\footnotesize #3}
\end{flushleft}
\end{table} }
\newcommand{\setab}[3]{
\end{tabular*}
\end{footnotesize}

\vspace*{#1}
\begin{flushleft}
\parbox{#2}{\footnotesize #3}
\end{flushleft}
\end{table} }

\def\eqan{\begin{equation}}
\def\eqen{\end{equation}}

\def\h1{\ion{H}{i}}
\def\h2{\ion{H}{ii}}
\def\reff{{\it R$_{\rm eff}$}}

\newcommand{\msun}{M$_\odot$}

\def\rr{{\sl R}$^{\star}$}
\def\P25{{\sl R}$_{\rm SF}$}
\def\E25{{\sl R}$_{\rm host}$}

\def\mstar{${\cal M}_{\star}$}

\def\m5{${\cal M}_{\star,{\rm 5\,Gyr}}$}

\def\D4000{$D_{4000}$}

\def\uflux{erg\,s$^{-1}$\,cm$^{-2}$}
\def\ha{H$\alpha$}
\def\ewha{EW(\ha)}

\def\e16{$10^{-16}~{\rm erg\,s^{-1}\,cm^{-2}}$}
\def\e17{$10^{-17}~{\rm erg\,s^{-1}\,cm^{-2}}$}
\newcommand{\PutLabel}[3]{\put(#1,#2){#3}}
\newcommand {\aga} {\ {\raise-.5ex\hbox{$\buildrel>\over\sim$}}\ }
\newcommand {\ala} {\ {\raise-.5ex\hbox{$\buildrel<\over\sim$}}\ } 

\begin{document}

% :::::::::::::::::::::::::::::::::::::::::::::::::::::::::::::::::::::::::::::::::
\titlerunning{Spectroscopic aperture biases in inside-out evolving early-type galaxies from CALIFA}

\authorrunning{J.~M.~Gomes et~al.}

\title{Spectroscopic aperture biases in inside-out evolving early-type galaxies from CALIFA
\thanks{Based on observations collected at the Centro Astron\'omico
Hispano Alem\'an (CAHA) at Calar Alto, operated jointly by the Max-Planck-Institut 
f\"ur Astronomie (MPIA) and the Instituto de Astrof\'isica de
Andaluc\'ia (CSIC)}}
% \subtitle{...}

\author{
J.~M.~Gomes\inst{\ref{inst1}}
\and P.~Papaderos\inst{\ref{inst1}}
\and J.~M.~V\'{\i}lchez\inst{\ref{inst2}}
\and C.~Kehrig\inst{\ref{inst2}}
\and J.~Iglesias-P\'aramo\inst{\ref{inst2},\ref{inst3}}
\and I.~Breda\inst{\ref{inst1}}
\and M.~D.~Lehnert\inst{\ref{inst4}}
\and S.~F.~S\'anchez\inst{\ref{inst5}}
\and B.~Ziegler\inst{\ref{inst6}}
\and S.~N.~dos Reis\inst{\ref{inst1}}
\and J.~Bland-Hawthorn\inst{\ref{inst7}}
\and L.~Galbany\inst{\ref{inst8},\ref{inst9}}
\and D.~J.~Bomans\inst{\ref{inst10},\ref{inst11}}
\and F.~F.~Rosales-Ortega\inst{\ref{inst12}}
\and C.~J.~Walcher\inst{\ref{inst13}}
\and R.~Garc{\'\i}a-Benito\inst{\ref{inst2}}
\and I.~M\'arquez\inst{\ref{inst2}}
\and A.~del Olmo\inst{\ref{inst2}}
\and M.~Moll\'a\inst{\ref{inst14}}
\and R.~A.~Marino\inst{\ref{inst15},\ref{inst16}}
\and C.~Catal\'an-Torrecilla\inst{\ref{inst17}}
\and R.~M.~Gonz\'alez Delgado\inst{\ref{inst2}}
\and \'A.~R.~L\'opez-S\'anchez\inst{\ref{inst18},\ref{inst19}}
\and the CALIFA collaboration
}
\offprints{jean@astro.up.pt}

\institute{
% Institute 1
Instituto de Astro{f\'i}sica e Ci{\^e}ncias do Espa\c{c}o, Universidade do Porto,
Centro de Astrof{\'\i}sica da Universidade do Porto, Rua das Estrelas, 4150-762 Porto, 
Portugal\label{inst1}
\and
% Institute 2
Instituto de Astrof\'isica de Andaluc\'ia (CSIC), Glorieta de la Astronom\'{\i}a s/n Aptdo. 3004, E18080-Granada, Spain\label{inst2}
\and
% Institute 3
Estaci\'on Experimental de Zonas Aridas (CSIC), Ctra. de Sacramento s.n., La Ca\~nada, Almería, Spain\label{inst3}
\and
% Institute 4
Institut d'Astrophysique de Paris, UMR 7095, CNRS, Universit\'{e} Pierre et Marie Curie, 98 bis boulevard Arago, 75014 Paris, France\label{inst4}
\and
% Institute 5
Instituto de Astronom\'ia,Universidad Nacional Auton\'oma de Mexico, A.P. 70-264, 04510, M\'exico, D.F.\label{inst5}
\and
% Institute 6
University of Vienna, T\"{u}rkenschanzstrasse 17, 1180 Vienna, Austria\label{inst6}
\and
% Institute 7
Sydney Institute for Astronomy, University of Sydney, NSW 2006,
Australia\label{inst7}
\and
% Institute 8
Millennium Institute of Astrophysics, Chile\label{inst8}
\and
% Institute 9
Departamento de Astronom\'ia, Universidad de Chile, Casilla 36-D, Santiago, Chile\label{inst9}
\and
% Institute 10
Astronomical Institute of the Ruhr-University Bochum\label{inst10}
\and
% Institute 11
RUB Research Department Plasmas with Complex Interactions\label{inst11}
\and
% Institute 12
Instituto Nacional de Astrof\'isica, \'Optica y Electr\'onica, Luis E. Erro 1, 72840 Tonantzintla, Puebla, Mexico\label{inst12}
\and
% Institute 13
Leibniz-Institut f\"ur Astrophysik Potsdam (AIP), An der Sternwarte 16, D-14482 Potsdam, Germany\label{inst13}
\and
% Institute 14
CIEMAT, Avda. Complutense 40, 28040 Madrid, Spain\label{inst14}
\and
% Institute 15
CEI Campus Moncloa, UCM-UPM, Departamento de Astrof\'isica y CC. de la
Atm\'osfera, Facultad de CC. F\'isicas, Universidad Complutense de Madrid,
Avda. Complutense s/n, 28040 Madrid, Spain\label{inst15}
\and
% Institute 16
Department of Physics, Institute for Astronomy, ETH Z\"urich, CH-8093
Z\"urich, Switzerland\label{inst16} 
\and
% Institute 17
Departamento de Astrof\'{\i}sica y CC. de la Atm\'{o}sfera, Universidad
Complutense de Madrid, E-28040, Madrid, Spain\label{inst17}
\and
% Institute 18
Australian Astronomical Observatory, PO Box 915, North Ryde,
NSW 1670, Australia\label{inst18}
\and
% Institute 19
Department of Physics and Astronomy, Macquarie University, NSW
2109, Australia\label{inst19}
} \date{Received ?? / Accepted ??}  

\abstract{Integral field spectroscopy (IFS) studies based on CALIFA survey
  data have recently revealed the presence of ongoing low-level star formation
  (SF) in the periphery of a small fraction ($\sim$10\%) of local early-type
  galaxies (ETGs), witnessing a still ongoing inside-out galaxy growth
  process.  A distinctive property of the nebular component in these ETGs,
  classified i+, is a two-radial-zone structure, with the inner zone
  displaying LINER emission with a \ha\ equivalent width \ewha\ $\simeq$ 1\AA,
  and the outer one (3\AA$<$\ewha$\la$20\AA) showing H{\sc ii}-region
  characteristics. Using CALIFA IFS data, we empirically demonstrate that the
  confinement of nebular emission to the galaxy periphery leads to a strong
  aperture (or, correspondingly, redshift) bias in spectroscopic 
single-fiber studies
  of type~i+ ETGs: At low redshift ($z \la 0.45$), SDSS spectroscopy is
  restricted to the inner (SF-devoid LINER) zone, thereby leading to their
  erroneous classification as `retired' galaxies, i.e. systems entirely
  lacking SF and whose faint nebular emission is solely powered by the
  post-AGB stellar component.  Only at higher $z$'s does the SDSS aperture
  progressively encompass the outer SF zone, permitting their unbiased
  classification as `composite SF/LINER'.  We also empirically demonstrate
  that the principal effect of a decreasing spectroscopic aperture on the
  classification of i+ ETGs via standard [N{\sc ii}]/H$\alpha$ vs [O{\sc
      iii}]/H$\beta$ emission-line (BPT) ratios consists in a monotonic
  up-right shift precisely along the upper-right wing of the `seagull'
  distribution on the BPT plane, i.e. the pathway connecting composite
  SF/H{\sc ii} galaxies with AGN/LINERs.
Motivated by these observational insights, we further investigate
theoretically observational biases in aperture-limited studies of inside-out
growing galaxies as a function of $z$. To this end, we devise a simple 1D
model, which involves an outwardly propagating, exponentially decreasing SF
process since $z \sim 10$ and reproduces the radial extent and two-zone
\ewha\ distribution of local i+ ETGs. By simulating on this model the
3\arcsec\ spectroscopic SDSS aperture, we find that SDSS studies at $z \la 1$
are progressively restricted to the inner (SF-devoid LINER) zone,
and miss an increasingly large portion of the \ha-emitting periphery.  This
leads to the false spectroscopic classification of such inside-out assembling
galaxies as retired ETG/LINERs besides a severe underestimation of their total
star formation rate (SFR) in a manner inversely related to $z$.  More
specifically, the SFR inferred from the \ha\ luminosity registered within the
SDSS fiber is reduced by 50\% at $z \sim 0.86$, reaching only 0.1\% of its
integral value at $z = 0.1$.  We argue that the aperture-driven biases
described above pertain to any morphological analog of i+ ETGs (e.g.,
SF-quiescent bulges within star-forming disks), regardless of whether it is
viewed from the perspective of inside-out growth or inside-out SF quenching,
and might be of considerable relevance to galaxy taxonomy and studies of the
cosmic SFR density as a function of $z$.}

\maketitle \markboth{Gomes et~al.}{Spectroscopic aperture biases
  in inside-out evolving early-type galaxies from CALIFA}

% ========================================================================
\section{Introduction \label{intro}}
% ========================================================================

Studies of large extragalactic samples with single-fiber spectroscopy from the
SDSS \citep[][]{yor00} and GAMA \citep{Driver09} have largely been relying on
the assumption that aperture-effects are negligible or can be accounted for in
a statistical sense. For example, \citet{Kewley2005} report that the condition
of the 3\arcsec\ SDSS fiber enclosing $\sim$20\% of the total emission of a
galaxy suffices for minimizing systematic and random aperture-related errors,
and recommend selecting samples at redshifts $z>0.04$.  More generally,
fundamental to numerous single-fiber studies
\citep[e.g.,][]{Kauffmann2003,Brinchmann2004,Tremonti2004} is the assumption
that the spectroscopic aperture encompasses a representative probe of
the spectrum of a galaxy, thus astrophysical quantities derivable from it
(e.g., the \ha\ luminosity) are either characteristic for a galaxy as a whole
or can be converted into integral ones through simple parametrizations.

% ::::::: Fig. 1 (integral spectrum and BPT diagnostics as a function of radius) :::::::::::::::::::::::::::::: (start)
\begin{figure*}
\begin{picture}(18.4,6.4)
\put(03.4,04.0){\includegraphics[width=2.4cm, viewport=20 10 520 290]{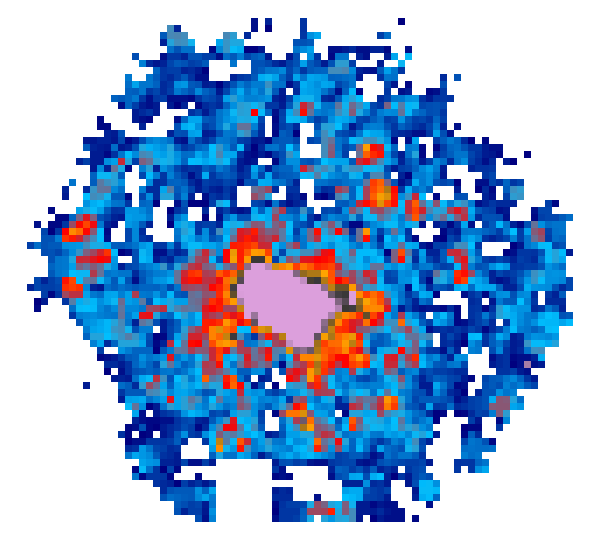}}
\put(01.1,04.0){\includegraphics[width=7.0cm, viewport=20 10 520 290]{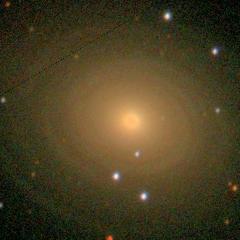}}
\put(09.3,03.9){\includegraphics[width=2.7cm, viewport=20 10 520 290]{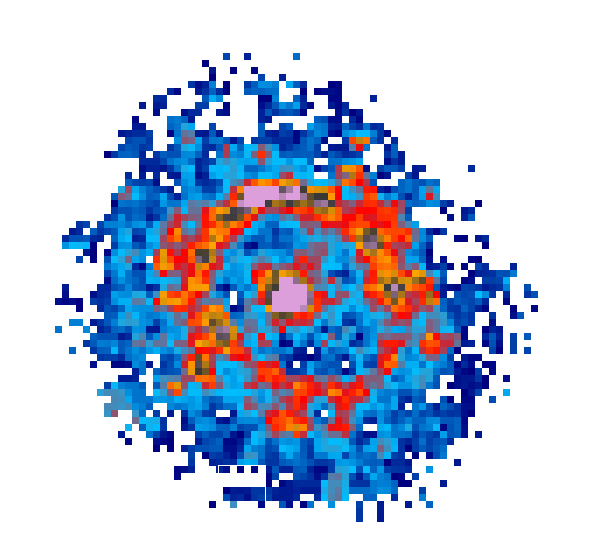}}
\put(07.1,04.0){\includegraphics[width=7.0cm, viewport=20 10 520 290]{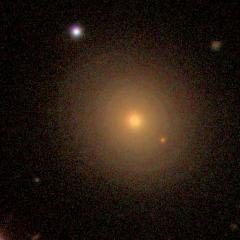}}
\put(15.5,03.7){\includegraphics[width=2.7cm, viewport=20 10 520 290]{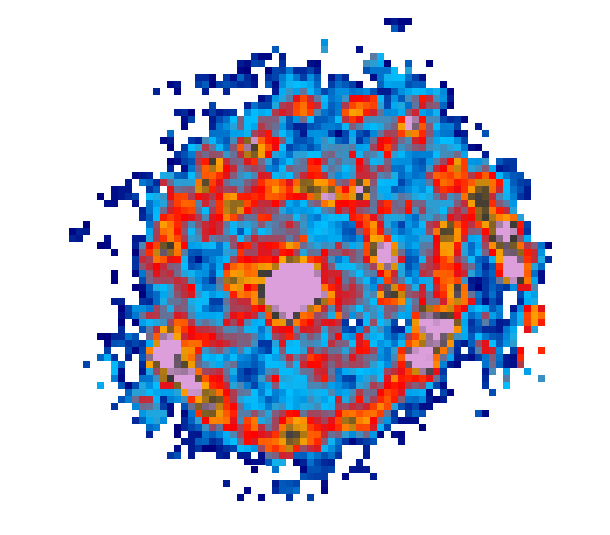}}
\put(13.4,04.0){\includegraphics[width=7.0cm, viewport=20 10 520 290]{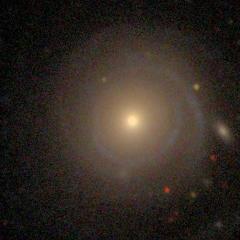}}
\put(00.0,-5.8){\includegraphics[width=6.8cm, viewport=20 10 520 290]{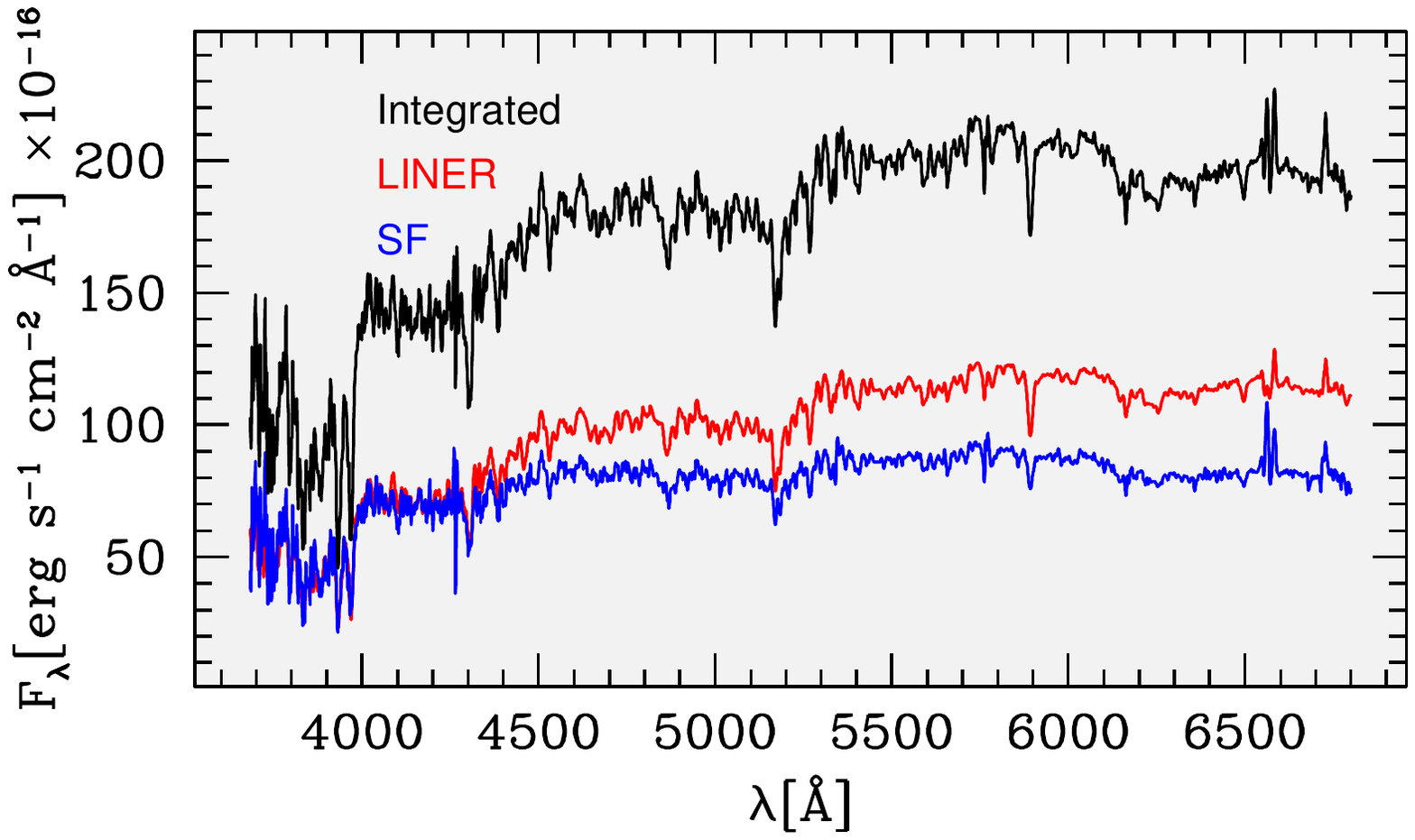}}
\put(06.0,-5.8){\includegraphics[width=6.8cm, viewport=20 10 520 290]{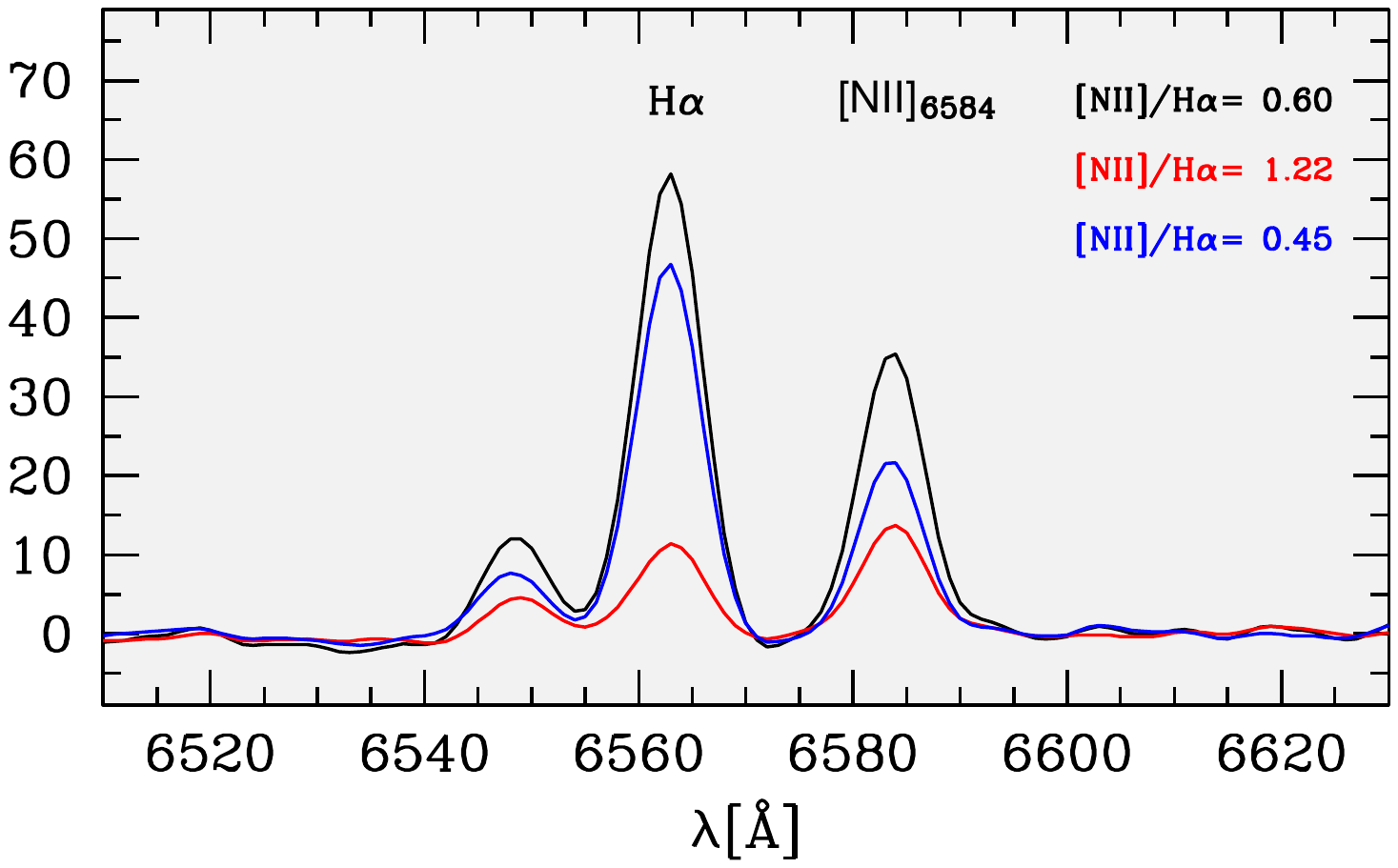}}
\put(12.3,-5.8){\includegraphics[width=6.8cm, viewport=20 10 520 290]{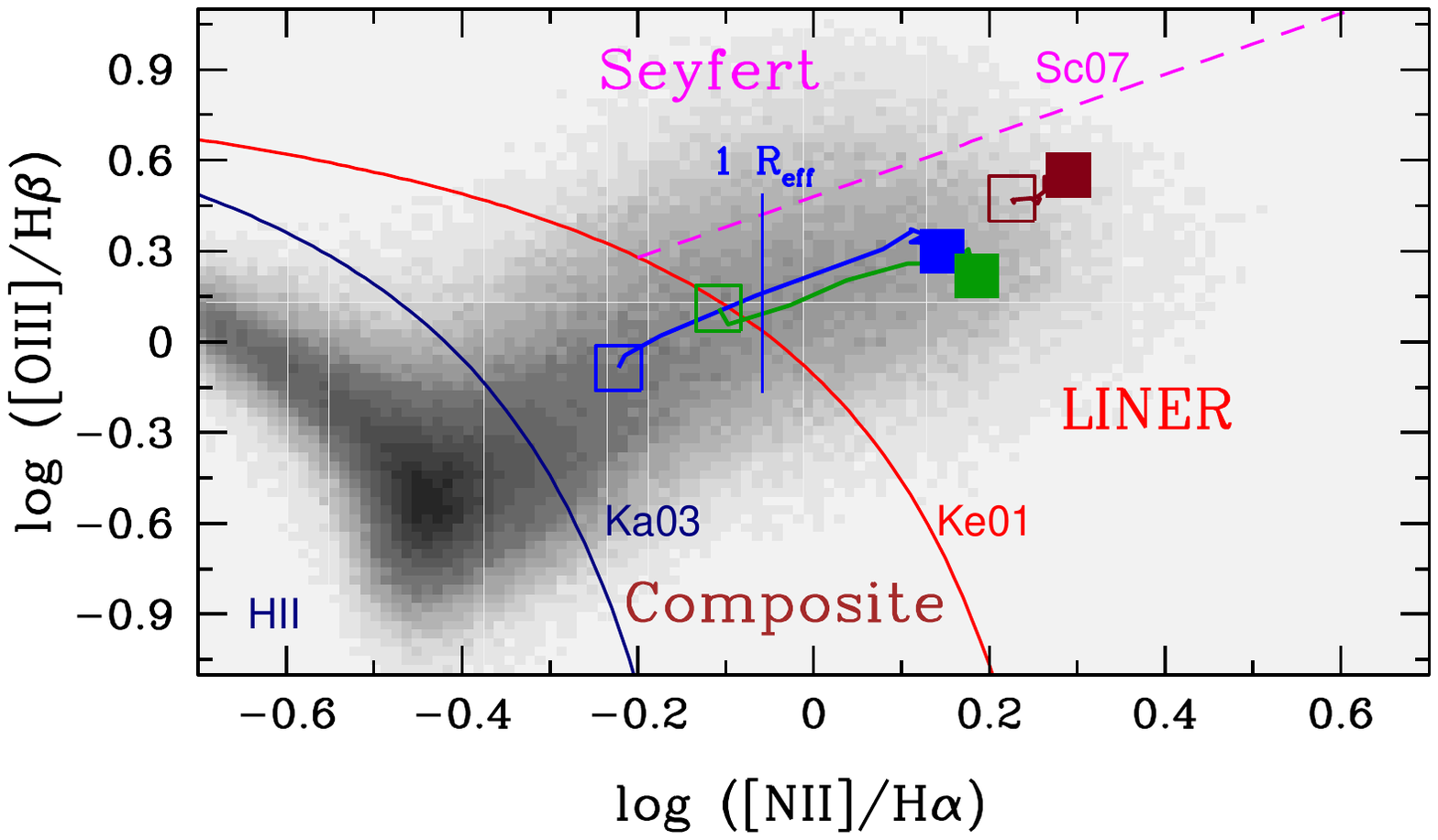}}
\PutLabel{00.90}{03.90}{\color{white}\vcap \object{NGC 1167}}
\PutLabel{06.90}{03.90}{\color{white}\vcap \object{NGC 1349}}
\PutLabel{13.20}{03.90}{\color{white}\vcap \object{NGC 3106}}
\PutLabel{01.00}{02.75}{\hvss a)}
\PutLabel{07.00}{02.75}{\hvss b)}
\PutLabel{13.30}{02.75}{\hvss c)}

% Halpha flux
\PutLabel{05.68}{06.12}{\vcap \ha}
\PutLabel{11.66}{06.12}{\vcap \ha}
\PutLabel{17.94}{06.12}{\vcap \ha}

\end{picture}
\caption[]{{\bf Upper panels:} SDSS true-color images and \ha\ maps (displayed
  between $0.05$ and $1 \times 10^{-16}$ \uflux) of the three i+ ETGs
  (\object{NGC 1167}, \object{NGC 1349} and \object{NGC 3106}; from left to
  right) studied in \cite{G15b,G15c}.  {\bf Lower panels: }{\bf a)} Comparison
  of the integrated spectrum of \object{NGC 1349} within its inner
  (\rr$\leq$8\arcsec) LINER zone (red color) with that obtained from its outer
  (\rr$>$8\arcsec) star-forming zone (blue color). The integrated spectrum of
  the ETG is overlaid in black color. {\bf b)} Zoom-in into the pure nebular
  spectrum of \object{NGC 1349} around the \ha\ 6563\AA\ Balmer line, as
  obtained after subtraction of the best-fitting stellar model to the observed
  spectrum (see P13 and G15a for details). The color coding is the same as in
  panel a.  Note the change of the \n2ha\ emission-line flux ratio from the
  inner (LINER) to the outer (SF-dominated) zone. {\bf c)} Variation of the
  diagnostic BPT ratios for the three i+ ETGs as a function of the aperture
  considered in their analysis. The brown, blue and green colors correspond to
  \object{NGC 1167}, \object{NGC 1349} and \object{NGC 3106}, respectively.
  The nuclear BPT ratios and those determined from the integrated galaxy
  spectra are shown with filled and open squares, respectively, and the
  connecting lines mark determinations based on successively larger
  spectroscopic apertures. The equivalent aperture radius in \reff\ is
  indicated for \object{NGC 1349} only for the sake of clarity. Note the shift
  of \object{NGC 1349} from the LINER into the `composite SF/LINER' regime of
  the BPT diagram when instead the inner zone ($\la$0.7\reff) the integral
  spectrum of the ETG is considered. The shaded background depicts the surface
  density of galaxies from SDSS in the upper-right wing of the BPT plane. The
  overlaid curves show the demarcation between AGN and LINERs
  \citep[][SC07]{Schawinski2007}, the locus of H{\sc ii} regions
  \citep[][Ka03]{Kauffmann2003}, and the `maximum SF' boundary
  \citep[][Ke01]{Kewley01}.}
\label{fig:spec}
\end{figure*}
% ::::::: Fig. 1 (integral spectrum and BPT diagnostics as a function of radius) :::::::::::::::::::::::::::::: (end)

This rationale permeates essentially the entire work that has dealt with large
extragalactic probes and their evolution with $z$ in the era of SDSS and GAMA,
and has gone a long way in our current understanding of a wide range of
topical issues related to, e.g., the dependence of the star formation rate
(SFR) and specific SFR on stellar mass (\mstar), the cosmic evolution of the
SFR density, and the spectral galaxy classification on the basis of diagnostic
emission-line ratios \citep[e.g., \tn2ha\ vs \to3hb,][hereafter BPT]{bpt81}.
For instance, determinations of integral SFRs had by necessity to rely on an
extrapolation of emission-line measurements within the spectroscopic aperture
assuming that the \ha\ scales with the underlying stellar continuum throughout
the galaxy's extent or is linked to broad-band colors.  For example,
\citet{Hopkins03} estimated integral SFRs from the \ha\ luminosity within the
spectroscopic aperture, and assuming a constant equivalent width (\ewha)
throughout the galaxy's extent. Similarly, \citet{Brinchmann2004} employed a
probabilistic estimate for the total SFR of SDSS galaxies that relates
broad-band colors with the \ha\ luminosity. An inherent weakness of these
approaches is obviously that, depending on the $z$ and the linear extent of a
galaxy, the fiber spectrum can be biased towards particular luminosity
entities (e.g., the brightest star-forming knot of a starburst galaxy or the
non-star-forming bulge of a late-type star-forming disk), making an
extrapolation to integral quantities uncertain.

% ::::::: Fig. 2 (integral spectrum and BPT diagnostics as a function of radius) :::::::::::::::::::::::::::::: (start)
\begin{figure*}
\begin{picture}(18.4,3.2)
\put(00.0,-5.9){\includegraphics[width=6.8cm, viewport=20 10 520 290]{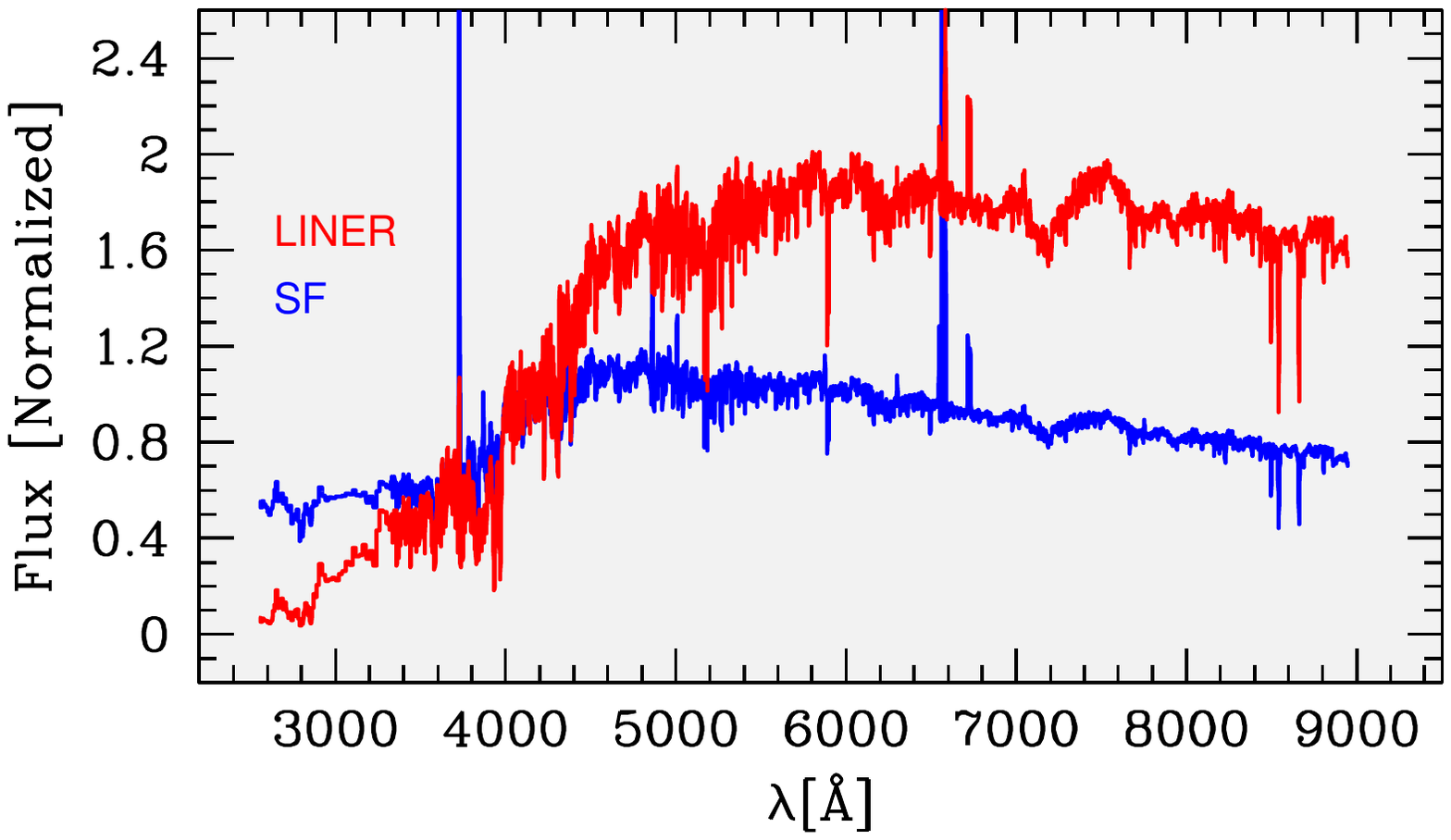}}
\put(06.0,-5.9){\includegraphics[width=6.8cm, viewport=20 10 520 290]{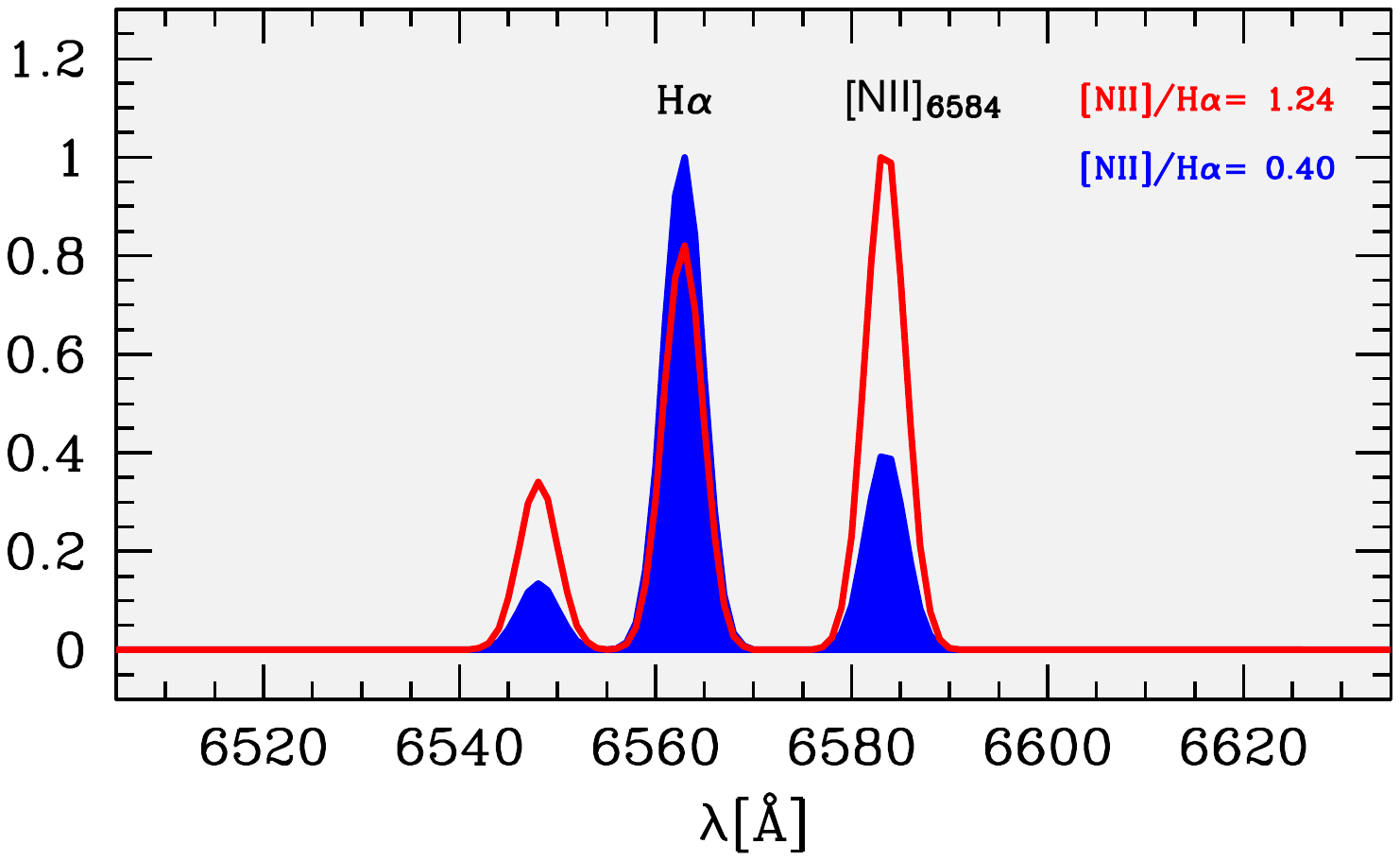}}
\put(12.3,-5.9){\includegraphics[width=6.8cm, viewport=20 10 520 290]{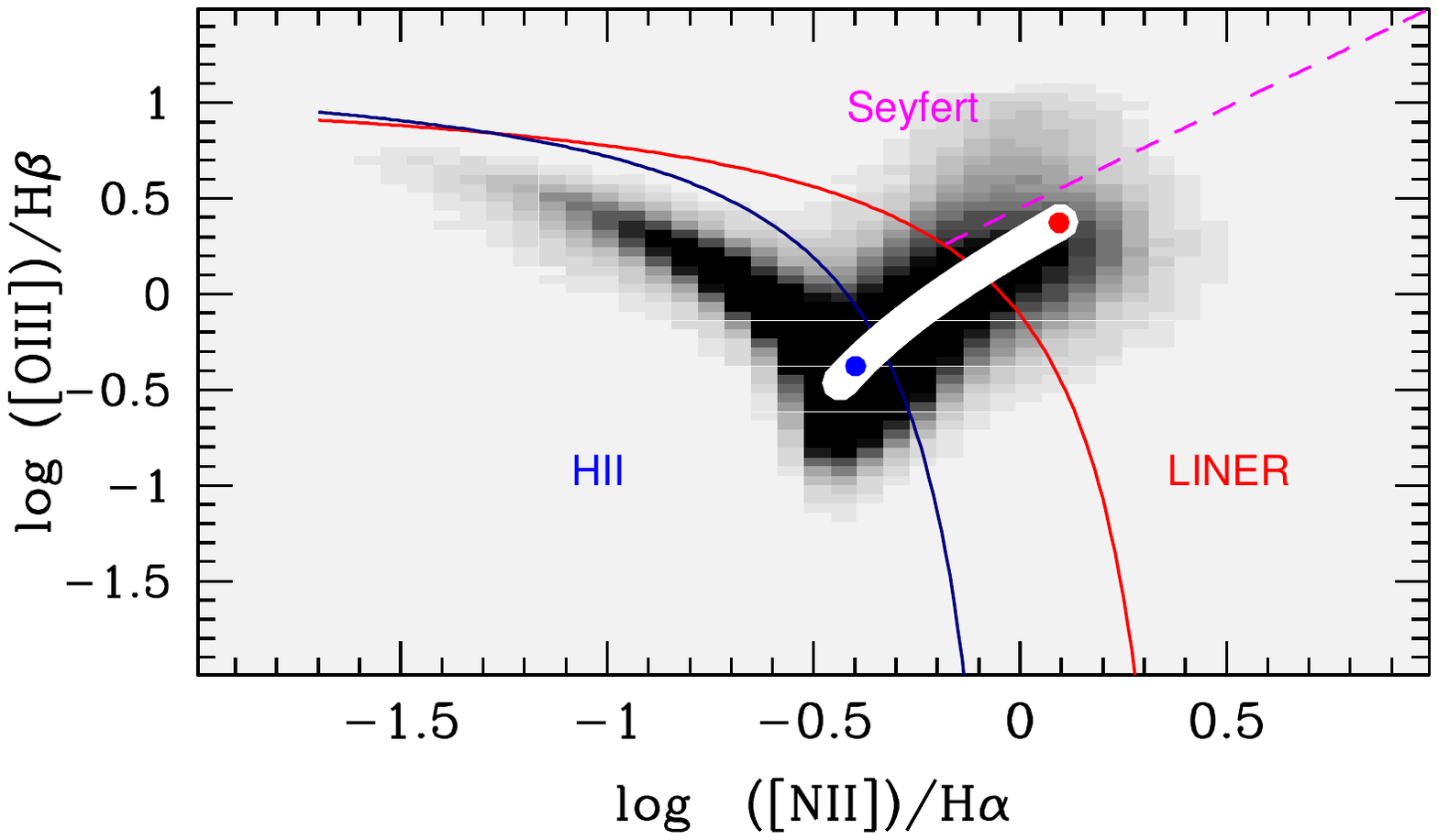}}
\PutLabel{01.00}{02.75}{\hvss a)}
\PutLabel{07.00}{02.75}{\hvss b)}
\PutLabel{13.30}{02.75}{\hvss c)}
\end{picture}

\caption[]{Simulation of spectroscopic aperture bias for an inside-out formed
  galaxy according to the 1D model described in Sect.\ref{model} (see
  Fig. \ref{fig:RadialProfiles}) that is observed with SDSS at a redshift $z =
  0.1$. At this $z$ the SDSS fiber (\diameter\ = 3\arcsec) covers a projected
  linear diameter of $\sim$5.5 kpc. {\bf a)} Comparison between the integrated
  spectrum of our model galaxy (blue color) with that obtained within the SDSS
  fiber (red color). Both spectra are normalized at 4020 \AA. {\bf b)}
  Zoom-in into the spectral region around the \ha\ line after removal of the
  underlying stellar continuum. The color coding is the same as in panel
  a. Note the considerable difference in the \n2ha\ emission-line ratio from
  the integral and simulated SDSS spectrum of the galaxy. From the former, the
  galaxy would be spectroscopically classifiable as a star-forming galaxy,
  while the latter would imply a purely retired LINER galaxy. {\bf c)}
  Variation of the BPT line ratios as a function of increasing aperture size
  for the adopted galaxy model (white color). It can be seen that that the
  simulated ratios form a continuous sequence along the right wing of
  the BPT diagram, moving from the LINER towards the H{\sc ii} zone as the
  aperture size increases, in close resemblance to the observational trend
  seen in Fig.~\ref{fig:spec}c.  The \n2ha\ and \o3hb\ ratios for the
  integrated and the simulated SDSS spectrum are shown in blue and red color,
  respectively. The grey shaded background depicts the surface density of
  galaxies from SDSS and the overlaid curves are the same as in
  Fig. {\ref{fig:spec}c}.}
\label{fig:Simulations}
\end{figure*}

The advent of integral field spectroscopy (IFS) over a large field of view
(FoV), as, e.g., from the Calar Alto Legacy Integral Field Area survey
\citep[CALIFA,][]{Sanchez2012}, has recently permitted empirical studies of
aperture-effects on various physical properties of the nebular emission in
late-type galaxies \citep[for instance, \ha\ luminosity and equivalent width
  EW; e.g.,][]{GMC12,IP13,Br13,Belfiore15}.

This subject has, however, not been investigated in similar detail for
early-type galaxies (ETGs), partly because of the faintness of their nebular
emission and the associated uncertainties in its study, and, presumably also,
due to the widespread view that these systems are comparatively simple and
spatially homogeneous in the characteristics of their stellar and nebular
components. This picture has now been substantially revised through IFS
studies, which continue to impressively reveal a great deal of complexity in
both \citep[e.g.,][hereafter
  G15a]{Sarzi06,McDermid07,Sarzi10,Kraj11,K12,Arnold13,Houghton13,Pracy14,G15a},
with conspicuous radial trends in stellar age
\citep[e.g.,][G15a]{Perez13,Rosa14} and the \ewha\ \citep[G15a,][hereafter
  P13]{P13} of local ETGs.  For a better understanding of aperture effects on
spectroscopic BPT classifications of ETGs it appears specially important to
investigate in a spatially resolved manner with IFS data the gas excitation
mechanisms in these systems.  Several studies over the past years have pointed
out the role of low-level SF activity \citep[see, e.g.,][and references
  therein]{Trager2000,Schawinski2007,Shapiro2010} on the excitation of faint
nebular emission in ETGs, as an alternative or supplementary mechanism to
photoionization by an active galactic nucleus \citep[AGN; e.g.,][]{Ho2008} or
the evolved ($\geq 10^8$ yr) post-AGB (pAGB) population
\citep[e.g.,][]{tri91,bin94,sta08}, or fast shocks
\citep[e.g.,][]{DopitaSutherland1995}.  Observational evidence for SF in ETGs
has been accumulating from multi-wavelength studies
\citep[e.g.,][]{Kaviraj07,GdP07,Schawinski2009,HuangGu09,Salim12,Petty13,Ko14,Pan14}.
For example, \citet{Kaviraj08} find that 10--15\% of the stellar mass (\mstar)
in these systems has been built in a declining SF process since $z \simeq 1$,
a conclusion that appears to be in line with the presence of a small fraction
($\sim$5.7\%) of blue ETGs in the local universe with estimated SF rates
(SFRs) between 0.5 and 50 \msun/yr \citep{Schawinski2009}.
Noteworthy in this regard is also 
that spatially resolved studies of individual ETGs indicate an
outwardly increasing luminosity contribution from young-to-intermediate-age
stellar populations \citep[e.g.,][]{Fang2012,G15a}.

In particular, \citet[][see also G15b]{G15a} identified a small fraction
($\sim$10\%) of CALIFA ETGs (classified as type i+) that show a steep
\ewha\ increase in their periphery. As demonstrated in a subsequent study
\citep[][hereafter G15c]{G15c} this outer \ewha\ excess is due to SF,
reflecting a still ongoing inside-out galaxy buildup process.

Central to our considerations in this study is the fact that the dominant
fraction (60--80\%) of the total \ha\ emission in all but one i+ ETGs studied
in G15c arises beyond one effective radius \reff\ ($\geq$10\arcsec), it thus
evades detection within the 3\arcsec\ SDSS aperture. Our aim here is to
explore aperture-effects on spectroscopic SDSS studies of type~i+ ETGs and
other inside-out forming galaxies as a function of cosmic time. To this end,
we take a twofold approach combining an empirical assessment of aperture
biases using as templates CALIFA IFS data for the galaxies studied in G15c
(Sect.~\ref{biases}) along with simulations on the basis of a simplified
inside-out galaxy growth model (Sect.~\ref{model}). Our conclusions are
summarized in Sect.~\ref{summary}.

% ============================================================================================================
\section{Anatomy of type~i+ ETGs and aperture biases on their spectroscopic classification\label{biases}}
% ============================================================================================================
The defining characteristics of i+ ETGs were discussed in G15a-c on the basis
of three CALIFA galaxies (Fig. \ref{fig:spec}). A distinctive property of the
nebular component of these systems is a two-radial-zone structure, with the
inner zone containing faint (\ewha$\simeq$1\AA) LINER emission, and the outer
one (3\AA$<$\ewha$\la$20\AA) essentially showing H{\sc ii}-region
characteristics.

This two-zone morphology can be best illustrated in the case of \object{NGC
  1349}: Its central part (\rr$\leq$8\arcsec, equivalent to 3.4 kpc) shows a
low, nearly constant \ewha\ of 1~\AA, which is consistent with gas
photoionization by the evolved ($\geq 10^8$ yr) post-AGB stellar component
(cf, e.g., G15a), whereas the high ($\sim$10~\AA) \ewha\ and BPT ratios in its
periphery (\rr$\geq$8\arcsec) imply gas photoionization by massive OB stars.
Using the isophotal annuli technique \citep[][P13]{P02} we extracted the
spectrum within the inner and outer zone, which are shown overlaid with the
integrated spectrum in Fig.~\ref{fig:spec}a. The middle panel shows the pure
nebular component in the region around the \ha\ Balmer line, as obtained after
subtraction of the best-fitting stellar model (see G15a for details).  From
this diagram it can be appreciated how significantly the \n2ha\ flux ratio
decreases from the inner to the outer zone. This effect can be better
evaluated from panel~c where we illustrate how the position of the type~i+
ETGs under study changes on the BPT plane when the galaxy spectrum is sampled
within successively larger annuli: In the case of \object{NGC 1167}, inclusion
of its very faint SF rim has a marginal impact on its AGN/LINER
classification, whereas in \object{NGC 1349} the much brighter peripheral SF
zone results in a down-left shift beneath the LINER demarcation curve, moving
the ETG into the locus of `composite SF/LINER' galaxies.  Had this galaxy been
observed within a FoV of $\diameter \leq$ 20\arcsec, then only its central,
almost emission-line free zone would have been considered in the analysis,
prompting its spectroscopic classification as a retired ETG/LINER, in the
definition by \citet{sta08}. Obviously, the same conclusion would had been
drawn from SDSS or GAMA spectra \citep{yor00,Baldry2010}. Only beyond $z\ga
0.45$ would the aperture of those surveys encompass the whole ETG, revealing
its true nature. Therefore, in the specific context of i+ ETGs (i.e. systems
where SF is mainly confined to the galaxy periphery), the decreasing
(increasing) proportion of star-forming (retired) galaxies with decreasing
redshift \citep[e.g.,][]{Stasinska15} could be partly driven by aperture
effects.

In summary, a decrease in the spectroscopic aperture can result in the case of
i+~ETGs in an up-right shift precisely along the right wing of the `seagull'
distribution on the BPT plane, i.e. the pathway connecting SF/H{\sc ii}
galaxies with AGN/LINERs.  This empirical fact calls for a critical
reconsideration of the way this right wing should be interpreted in addressing
the relative role of thermal and non-thermal activity in i+ ETGs and their
morphological analogs (e.g., late-type galaxies with an old, SF-devoid bulge
within a more extended star-forming disk).

% ============================================================================================================
\section{Inside-out galaxy formation and associated aperture effects
\label{model}}
% ============================================================================================================
\begin{figure}
\includegraphics[width=10.8cm,height=4.0cm, viewport=30 430 720 690]{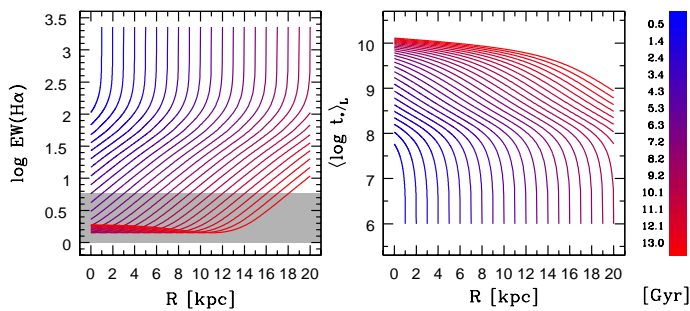}
\caption[]{ Radial distribution of the \ewha\ and the light-weighted mean
  stellar age ($\langle \log t_\star \rangle_L$), as predicted by the adopted
  inside-out galaxy formation scenario for 28 ages between 0 and 13.52 Gyr (cf
  color coding on the right-side bar).The region where the observed \ewha\ is
  consistent with pure pAGB photoionization ($<2.4$\AA) is shown with the
  shaded grey area (see G15a). The model adopts a constant age gradient
  $\nabla t = -0.5$ Gyr/kpc and outwardly propagating SF at $v\sim$2 km/s. In
  this particular simulation when the galaxy reaches a radius of 20 kpc the
  inside-out growth ceases. The observational properties are computed for
  each zone assuming an exponentially declining SFR with an e-folding
  timescale of 1 Gyr.}
\label{fig:RadialProfiles}
\end{figure}

% ::::::: Fig. X SDSS aperture
\begin{figure}[!]
\includegraphics[width=12.0cm, viewport=40 250 750 690]{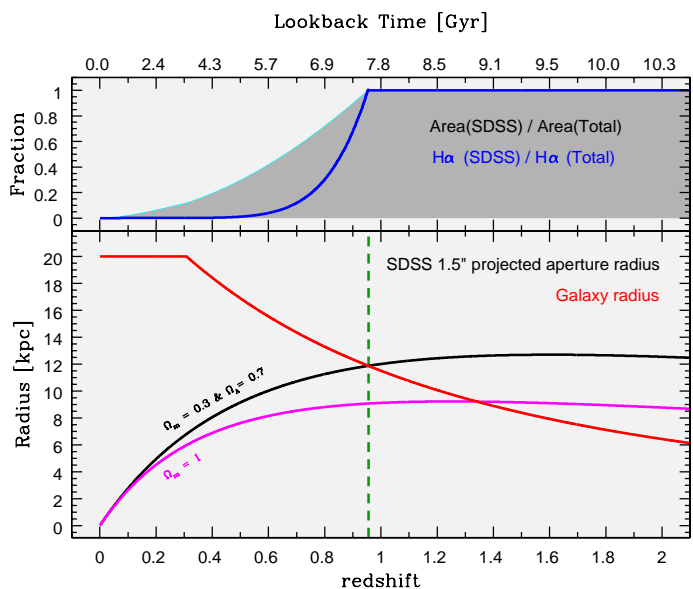}
\caption[]{{\bf Top panel:} Fraction of the area (shaded region) and of the
  total \ha\ flux (blue curve) encompassed by the 3\arcsec\ SDSS fiber as a
  function of $z$ (or, equivalently, lookback time in Gyr; upper label). {\bf
    Bottom panel:} Radius projected within the SDSS fiber as a function of $z$
  for a Friedmann-Robertson-Walker cosmology with H$_0 = 70$ km s$^ {-1}$
  Mpc$^{-1}$, $\Omega_m = 0.3$ and $\Omega_\Lambda = 0.7$ (black curve) and,
  for the sake of comparison, a flat, matter-dominated Universe ($\Omega_m =
  1$; magenta curve). The radius of the inside-out forming galaxy is drawn in
  red and the dashed line marks the $z=0.95$ below which the area subtended by
  the SDSS fiber becomes smaller than the galaxy. }
\label{fig:SDSSAperture}
\end{figure}

Motivated by the observational evidence laid out in G15c and in the previous
section, we extend our study by considering an inside-out formation process
\citep[e.g.,][]{Perez13,Rosa14} for ETGs. This is simulated by an outwardly
propagating SF process in a 1D galaxy model, which, combined with an
evolutionary synthesis code (cf G15a), permits computation of the time
evolution of various spectrophotometric properties as a function of radius.
 
A SF propagation process with a velocity $v$ can be written by making use of
the 1D wave equation:

\begin{equation}
\nabla^2 {\rm SFR}(r,t) = \frac{\partial^2 {\rm SFR}(r,t)}{\partial r^2} = \frac{1}{v^2} \frac{\partial^2{\rm
    SFR}(r,t)}{\partial t^2}
\end{equation}

\noindent where the solution of this partial differential equation is given by
\citet{dalembert1747} with a linear sum of two arbitrary functions $f(v t -
r)$ and $g(v t + r)$ that represent an incoming and outgoing wave,
respectively.  Since the SFR wave is propagated outwardly due to the
inside-out growth of the galaxy, we can write the general solution as
${\rm SFR}(r,t) = f(v t - r) = \phi(t - r / v)$.

For each radial zone, the SFR$(r,t)$ is chosen to be exponentially declining
as $\propto e^{-(t-r/v)/\tau}$ with an e-folding timescale of $\tau = $1 Gyr,
constant wave velocity $v$ and subject to the constraint $t \geq r/v$. This
scenario is compatible with a quick cessation of SF in the inner zone (as
expected, e.g., during the bulge formation) and continued residual SF in the
galaxy periphery in an inside-out galaxy buildup process. For simplicity, the
galaxy is constructed such as to display a constant age gradient $\nabla
t=-0.5$ Gyr/kpc, which imposes a maximum radius as a function of lookback time
of $R_{\rm max}(t) = (t_0-t) / |\nabla t|$, where $t_0$ is the galaxy
formation lookback time. For the adopted model, the galaxy grows radially at a
constant speed $v \sim 2 {\rm km/s}$ and when attains $R_{\rm max}=20$ kpc (at
$z \sim 0.3$) the linear growth ceases, yielding a typical present-day ETG
radius \citep[e.g.,][]{G15a}. The Friedmann-Robertson-Walker cosmology with
H$_0 = 70$ km s$^ {-1}$ Mpc$^{-1}$, $\Omega_m = 0.3$ and $\Omega_\Lambda =
0.7$ has been adopted, yielding 13.52 Gyr for the age for the Universe. The
time $t_0$ of galaxy formation is taken to be $\sim$ 0.5 Gyr (redshift 10)
after the Big Bang.

The spectral energy distribution (SED) is computed following the detailed
prescriptions by G15a. The synthetic composite stellar SEDs were evaluated
using the full set of ages for the Simple Stellar Populations (SSPs) from
\citet[][hereafter BC03]{bru03}; comprising $\sim$200 spectra spanning an age
between 0 and 13.52 Gyr, and assuming a constant solar metallicity.  Note
that, even though emission-line luminosities and their ratios for sub-solar
metallicities can significantly change due to the harder, increased UV
ionizing output of massive low-metallicity stars, the principal trends and
conclusions inferred from the adopted inside-out galaxy growth model remain
unaltered. The BC03 SSP library uses `Padova 1994' evolutionary tracks
\citep{Alongi1993,Bressan1993,Fagotto1994a,Fagotto1994b,Fagotto1994c,Girardi1996},
the \citet{Chabrier2003} initial mass function and the STELIB stellar library
\citep{LeBorgne2003}.

Hydrogen (Balmer, Paschen, etc...) line fluxes were computed from the total UV
ionizing flux assuming case~B recombination ($T_{\rm e} = 10000$ K and $n_{\rm
  e} = 100$ cm$^{-3}$) with the corresponding effective recombination
coefficient $\alpha^{eff}_{{\rm H}\alpha}$ and by assuming ionization-bound
nebulae. The fluxes of collisionally excited lines are based on semi-empirical
calibrations for a) classical H{\sc ii} regions \citep[e.g.,][]{Anders2003},
photoionized by OB stars and b) ETG nuclei, photoionized by post-AGB stars
(old stellar populations $\geq 1$Myr) and showing LINER BPT ratios
\citep[e.g.,][]{bin94}. Therefore, the total UV ionizing flux coming from an
exponentially declining SFR model with $\tau = 1$ Gyr contains the combined
output from a young SF plus an old post-AGB component in certain evolutionary
phases that produces a luminosity-weighted average of the modeled
emission-lines.

The computed radial profiles for the \ewha\ and light-weighted stellar age for
various evolutionary stages (from 0 to 13.52 Gyr) are shown in
Fig. \ref{fig:RadialProfiles}. It is apparent that the \ewha, in general,
decreases as a function of time while the second always increases.

The 3\arcsec\ SDSS aperture is subsequently simulated on the model galaxy in
order to evaluate aperture effects and their dependence on $z$. In the adopted
cosmological model, the linear radius projected within the SDSS fiber as a
function of $z$ was simulated on the model (Fig. \ref{fig:SDSSAperture}).  It
can be seen that for $z \la 1$ the SDSS aperture encompasses a progressively
smaller fraction of the galaxy. Quite importantly, the fraction of
\ha\ luminosity registered within the SDSS fiber (upper panel) has a much
steeper decline than the fractional area covered by SDSS, due to the
confinement of SF activity to the galaxy periphery. Quantitatively, the
\ha\ flux registered within the SDSS fiber decreases by 50\% at $z\sim 0.86$,
reaching only 0.1\% of its integral value at $z = 0.1$.

Consequently, a strong aperture bias is to be expected in SFR determinations
and the spectroscopic classification of inside-out assembling i+ ETGs, and
their morphological analogs. In order to exemplify this bias, a snapshot at
$z\sim 0.1$ is shown in Fig.~\ref{fig:Simulations}. At this stage, in which
the model galaxy has reached its maximum radius, the SDSS fiber
samples a projected radius of $\sim$2.75 kpc ($\sim$2\% of the area of the
galaxy). As apparent from panel~a, the integrated galaxy spectrum is very
different than the one registered within the SDSS fiber: The former is
characteristic of a blue star-forming galaxy, showing BPT ratios typical of
H{\sc ii} regions, whereas the latter indicates a retired LINER/ETG.

This is also reflected on the BPT diagram (panel~c), where the white
curve delineates the variation of the BPT line-ratios obtained for our model
galaxy within a set of increasing apertures. The \n2ha\ and \o3hb\ ratios
corresponding to the integrated spectrum and those registered within the SDSS
aperture are shown as blue $\left(0.40,0.42\right)$ and red $(1.24,2.36)$
dots, respectively. It can be seen that the simulated BPT ratios describe a
continuous sequence along the right wing of SDSS determinations (shaded
area), connecting the LINER with the H{\sc ii}/SF zone of the BPT parameter
space, in agreement with the observational trend described in
Sect. \ref{biases} (Fig.~\ref{fig:spec}c). Specially important in this
context is that the identical trend arises when, instead of using successively
larger apertures at $z=0.1$, the SDSS aperture is simulated on the inside-out
evolving galaxy across $z$.

In summary, for an inside-out assembling galaxy with the model assumption
adopted here, the simultaneous linear and angular growth of the inner (SF
devoid/LINER) zone will cause limited-aperture surveys to strongly
underestimate the total SFR in a manner inversely related to $z$. This bias,
arising at $z \la 1$ and becoming progressively severe for lower $z$'s could
artificially increase the estimated fraction of non-star-forming (retired)
galaxies over the past $\sim$7 Gyr, impacting galaxy classification and
mimicking a steeper decline of the cosmic SFR density.

% ==========================================================================
\section{Summary and conclusions \label{summary}}
% ==========================================================================
This study has been motivated by the recent detection by \citet{G15a,G15b} of faint
spiral-like features in the low-surface brightness periphery of a small subset
($\sim$10\%) of nearby early-type galaxies (ETGs) from the CALIFA integral
field spectroscopic (IFS) galaxy survey. As a subsequent analysis by
\citet{G15c} has revealed, these features witness a still ongoing
inside-out galaxy growth process and are spatially associated with an extended
emission-line zone containing up to $\sim$80\% of the total \ha\ emission of
these galaxies.  A distinctive property of the nebular component in these
ETGs, classified as i+, is a two-radial-zone structure, with the inner
zone being of the order of the galaxy's effective radius
\reff\ ($\sim$10\arcsec) and containing faint (\ewha$\simeq$1~\AA) LINER
emission, and the outer one (3\AA$<$\ewha$\la$20\AA) that displays H{\sc
  ii}-region characteristics.
  
A question naturally arising by the observed segregation of the nebular
emission in i+ ETGs in two spatially and physically distinct concentric zones
is, how aperture effects might impact spectroscopic studies of these systems.
To address this issue, we take here a combined empirical and theoretical
approach, aiming at a qualitative assessment of aperture-driven biases in SDSS
studies of type~i+ ETGs and other inside-out forming galaxies across cosmic
time.

At a first stage, using CALIFA IFS data, we empirically demonstrate that, for
a typical i+ ETG, the confinement of nebular emission to the galaxy periphery
leads to a strong observational bias in spectroscopic studies with SDSS. At
low redshift ($z \la 0.45$), the 3\arcsec\ SDSS fiber captures only the inner
(non-star-forming LINER) zone of such a galaxy, leading to its erroneous
classification as retired, i.e. a system entirely lacking ongoing star
formation (SF) and whose faint nebular emission is solely powered by the
post-AGB stellar component. Only at higher $z$'s does the SDSS aperture
progressively encompass the outer (star-forming) zone permitting its unbiased
classification as `composite SF/LINER'.  We also empirically demonstrate that
the principal effect of a decreasing spectroscopic aperture on the
classification of i+ ETGs via standard [N{\sc ii}]/H$\alpha$ vs [O{\sc
    iii}]/H$\beta$ emission-line (BPT) ratios consists in a monotonic up-right
shift precisely along the right-wing of the `seagull' distribution on
the BPT plane, i.e. the pathway connecting composite SF/H{\sc ii}
galaxies with AGN/LINERs.

These empirical insights are further underscored through a simple 1D
inside-out galaxy formation model involving an outwardly propagating,
exponentially decreasing SF process since $z=10$, which reproduces both
the radial extent and two-zone \ewha\ morphology of present-day i+ ETGs.  By
simulating on this model the SDSS aperture, we find that, for $z \la 1$,
SDSS spectroscopy is progressively restricted to the inner (SF-devoid LINER)
zone of an inside-out forming galaxy, thereby missing an increasingly large
portion of its \ha-emitting periphery.  More specifically, according to our
model, the \ha\ flux registered within the SDSS fiber decreases by 50\% at
$z\simeq 0.86$, reaching only 0.1\% of its integral value at $z=0.1$.

Our model also reproduces for local ($z\simeq0.1$) i+ ETGs the observed
variation of BPT line-ratios with aperture size along the right-wing of SDSS
determinations, lending further support to the conjecture that spectroscopic
classification on the basis of SDSS data is prone to substantial aperture
effects.  Particularly important in this context is that the same trend along
the right-wing is reproducible when, instead of using successively larger
apertures for a local model-ETG, the SDSS aperture is simulated on an
inside-out forming galaxy since $z=10$ ($\sim$13 Gyr).

The combined empirical and theoretical evidence from this study therefore
suggests that the right-wing distribution of BPT determinations for i+ ETGs
and their morphological analogs (e.g., late-type galaxies with an old,
SF-devoid bulge centered on a more extended star-forming disk) with
single-fiber spectroscopy (e.g., SDSS, GAMA) is consistent with (yet no proof
for) a pure aperture effect, and naturally reproducible in an inside-out
galaxy growth scenario. This calls for a closer examination of the way this
right-wing should be interpreted in addressing the relative role of thermal
and non-thermal activity in these systems.

Finally, in the framework of the adopted inside-out galaxy formation model,
the simultaneous linear and angular growth of the inner (SF devoid/LINER) zone
will cause limited-aperture surveys (e.g., SDSS, GAMA) to strongly
underestimate the total star formation rate (SFR) in a manner inversely
related to $z$. This bias, arising at $z \la 1$ and becoming progressively
severe for lower $z$'s could artificially increase the estimated fraction of
non-star-forming (retired) galaxies over the past $\sim$7 Gyr, impacting
galaxy classification and mimicking a steeper decline of the cosmic SFR
density.

Such considerations underscore the critical importance of IFS studies of ETGs
(and galaxies in general) over their entire optical extent.

\begin{acknowledgements}
  This paper is based on data from the Calar Alto Legacy Integral Field Area
  Survey, CALIFA (http://califa.caha.es), funded by the Spanish Ministery of
  Science under grant ICTS-2009-10, and the Centro Astron\'omico
  Hispano-Alem\'an. JMG acknowledges support by Funda\c{c}\~{a}o para a
  Ci\^{e}ncia e a Tecnologia (FCT) through the Fellowship SFRH/BPD/66958/2009
  and POPH/FSE (EC) by FEDER funding through the program Programa Operacional
  de Factores de Competitividade (COMPETE). PP is supported by FCT through the
  Investigador FCT Contract No. IF/01220/2013 and POPH/FSE (EC) by FEDER
  funding through the program COMPETE. JMG\&PP also acknowledge support by FCT
  under project FCOMP-01-0124-FEDER-029170 (Reference FCT
  PTDC/FIS-AST/3214/2012), funded by FCT-MEC (PIDDAC) and FEDER (COMPETE). SFS
  acknowledges support from CONACyT-180125 and PAPIIT-IA100815 grants. Support
  for LG is provided by the Ministry of Economy, Development, and Tourism's
  Millennium Science Initiative through grant IC120009, awarded to The
  Millennium Institute of Astrophysics, MAS. LG acknowledges support by
  CONICYT through FONDECYT grant 3140566. CJW acknowledges support through the
  Marie Curie Career Integration Grant 303912. IM acknowledges financial
  support by the Junta de Andaluc\'ia through project TIC114, and the Spanish
  Ministry of Economy and Competitiveness (MINECO) through projects
  AYA2010-15169 and AYA2013-42227-P. R.A. Marino is funded by the Spanish
  program of International Campus of Excellence Moncloa (CEI). This research
  made use of the NASA/IPAC Extragalactic Database (NED) which is operated by
  the Jet Propulsion Laboratory, California Institute of Technology, under
  contract with the National Aeronautics and Space Administration.
\end{acknowledgements}

\end{document}